\documentclass[10pt,conference]{IEEEtran}
\usepackage{color}
\usepackage{multirow}
\usepackage{pifont}
\usepackage{afterpage}
\usepackage{rotating}
\usepackage{array}
\usepackage{amsmath}
\usepackage{txfonts}
\newcolumntype{x}[1]{>{\raggedright\arraybackslash}p{#1}}

\usepackage{makecell}
\usepackage{graphicx}
\usepackage{caption}
\usepackage{subcaption}
\usepackage{url}
\usepackage{booktabs}
\usepackage{tabularx}
\newcolumntype{L}[1]{>{\raggedright\let\newline\\\arraybackslash}p{#1}}  

\usepackage[
   pass,
]{geometry}

\begin{document}

\title{Semantically Enhanced Software Traceability\\Using Deep Learning Techniques} 

\author{
\IEEEauthorblockN{Jin Guo, Jinghui Cheng and Jane Cleland-Huang}
\IEEEauthorblockA{University of Notre Dame\\
Notre Dame, IN, USA\\
Email: \{jguo3, JinghuiCheng, JaneClelandHuang\}@nd.edu}
}

\maketitle
\begin{abstract}
In most safety-critical domains the need for traceability is prescribed by certifying bodies. Trace links are generally created among requirements, design, source code, test cases and other artifacts; however, creating such links manually is time consuming and error prone. Automated solutions use information retrieval and machine learning techniques to generate trace links; however, current techniques fail to understand semantics of the software artifacts or to integrate domain knowledge into the tracing process and therefore tend to deliver imprecise and inaccurate results. In this paper, we present a solution that uses deep learning to incorporate requirements artifact semantics and domain knowledge into the tracing solution. We propose a tracing network architecture that utilizes Word Embedding and Recurrent Neural Network (RNN) models to generate trace links. Word embedding learns word vectors that represent knowledge of the domain corpus and RNN uses these word vectors to learn the sentence semantics of requirements artifacts. We trained 360 different configurations of the tracing network using existing trace links in the Positive Train Control domain and identified the Bidirectional Gated Recurrent Unit (BI-GRU) as the best model for the tracing task. BI-GRU significantly out-performed state-of-the-art tracing methods including the Vector Space Model and Latent Semantic Indexing.
\end{abstract}

\begin{IEEEkeywords}
	Traceability, Deep Learning, Recurrent Neural Network, Semantic Representation.
\end{IEEEkeywords}

\section{Introduction}
\label{sec:intro}
Requirements traceability plays an essential role in the software development process.  Defined as ``the ability to describe and follow the life of a requirement in both a forwards and backwards direction through periods of ongoing refinement and iteration'' \cite{DBLP:conf/re/GotelF94}, traceability supports a diverse set of software engineering activities including change impact analysis, regression test selection, cost prediction, and compliance verification \cite{DBLP:books/daglib/p/GotelCHZEGDAMM12}. In high-dependability systems, regulatory standards, such as the US Federal Aviation Authority's (FAA) DO178b/c \cite{DO178b}, prescribe the need for trace links to be established and maintained between hazards, faults, requirements, design, code, and test cases in order to demonstrate that a system is safe for use  \cite{DBLP:conf/icse/Knight02, FDA}. Unfortunately, the tracing task is arduous to perform and error-prone \cite{DBLP:conf/iwpc/MahmoudNX12}, even when industrial tools are used to manually create links or to capture them as a byproduct of the development process \cite{DBLP:conf/icse/Cleland-HuangGHMZ14}. In practice, trace links are often  incomplete and inaccurate \cite{DBLP:conf/sigsoft/Cleland-HuangRM14}, even in safety-critical systems \cite{DBLP:journals/software/MaderJZC13, DBLP:conf/se/RempelMKC15}. 

To address these problems, researchers have proposed and developed solutions for automating the task of creating and maintaining trace links \cite{Antoniol:Recovering,DeLucia:ArtefManag,DBLP:journals/tse/HayesDS06}. Solutions have included information retrieval approaches \cite{DeLucia:ArtefManag, DBLP:conf/re/DekhtyarHSHD07, DBLP:conf/icse/AsuncionAT10}, machine learning \cite{DBLP:journals/re/MahmoudW16,DBLP:conf/msr/0004RCRHV16, DBLP:conf/msr/0004RCRHV16, DBLP:conf/re/NiuM12}, heuristic techniques \cite{DBLP:journals/jss/SpanoudakisZPK04, DBLP:conf/re/0004CB13}, and AI swarming algorithms \cite{DBLP:journals/re/SultanovHK11}.  Other approaches, especially in the area of feature location \cite{DIT/EMSE/2012}, require additional information obtained from runtime execution traces. Results have been mixed, especially when applied to industrial-sized datasets, where acceptable recall levels above 90\% can often only be achieved at extremely low levels of precision \cite{DBLP:conf/sigsoft/LoharAZC13}.  

One of the primary reasons that automated approaches have underperformed is the term mismatch that often exists between pairs of related artifacts \cite{biggerstaff1994program}.  To illustrate this we draw on an example from the Positive Train Control (PTC) domain. PTC is a communication-based train control system designed to ensure that trains follow directives in order to prevent  accidents from occurring \cite{PTC}. The requirement stating that  \textit{``The BOS Administrative Toolset shall allow the Authorized Administrator to view an On-board's last reported On-board Software Version, including the associated repository name, MD5, and whether the fileset is preferred or acceptable.''} is associated with the design artifact stating that \textit{``The Operational Data Panel is used to provide information about the current PTC operations in a subdivision''}. Recognizing and establishing this link requires non-trivial knowledge of domain concepts -- for example, understanding that BOS Administrative Toolset contains the Operational Data Panel, each locomotive contains an On-board unit for PTC operation, and that the Operational Data Panel displays the information of locomotives such as the On-board Software Version to the BOS Authorized Administrator. This link would likely be missed by popular trace retrieval algorithms such as the Vector Space Model (VSM), Latent Semantic Indexing (LSI), and Latent Direchlet Allocation (LDA), which all represent artifacts as bags of words and therefore lose the artifacts' embedded semantics. It would also be missed by techniques that incorporate phrasing without understanding their conceptual associations \cite{DBLP:conf/re/0002SGBZ15, RAISE2014}.  In fact, most current techniques lack the sophistication needed to reason about semantic associations between artifacts and therefore fail to establish trace links when there is little meaningful overlap in use of terms. 

In our prior work we developed \emph{Domain-Contextualized Intelligent Traceability} (DoCIT) \cite{guo2014DoCIT} as a proof of concept solution to investigate the integration of domain knowledge into the tracing process.  We demonstrated that for the domain of PTC systems DoCIT returned accurate trace links achieving mean average precision (MAP) of .822 in comparison to .590 achieved using VSM.  However, the cost of setting up DoCIT for a domain was non-trivial, as it required carefully handcrafting a domain ontology and manually defining trace link heuristics capable of reasoning over the semantics of the artifacts and associated domain knowledge.  Furthermore, DoCIT depended upon a conventional syntactic parser to analyze the artifacts in order to extract meaningful concepts.  The approach was therefore sensitive to errors in the parser, terms missing from the ontology, and missing or inadequate heuristics.  As such, DoCIT was effective but fragile, and would require significant effort to transfer into new project domains.

On the other hand, deep learning techniques have successfully been applied to solve many Natural Language Processing (NLP) tasks including parsing \cite{socher2011parsing}, sentiment analysis \cite{tai2015tree_lstm}, question answering \cite{iyyer2014neural}, and machine translation \cite{DBLP:journals/corr/BahdanauCB14}.   Such techniques abstract problems into multiple layers of nonlinear processing nodes; they leverage either supervised or unsupervised learning techniques to automatically learn a representation of the language and then use this representation to perform complex NLP tasks. The goal of the work described in this paper is to utilize deep learning to deliver a scalable, portable, and fully automated solution for bridging the semantic gap that currently inhibits the success of trace link creation algorithms. Our solution is  designed to automate the capture of domain knowledge and the artifacts' textual semantics with the explicit goal of improving accuracy of the trace link generation task. 

The approach we propose includes two primary phases.  First, we learn a set of \textbf{word embeddings} for the domain using an unsupervised learning approach trained over a large set of domain documents. The approach generates high dimensional word vectors that capture distributional semantics and co-occurrence statistics for each word \cite{pennington2014glove}. Second, we use an existing training set of validated trace links from the domain to train a Tracing Network to predict the likelihood of a trace link existing between two software artifacts. Within the tracing network, we adopt a {\bf Recurrent Neural Network} architecture to learn the representation of artifact semantics. For each artifact (i.e. each regulation, requirement, or source code file etc.), each word is replaced by its associated vector representation learned in the word embedding training phase and then sequentially fed into the RNN. The final output of RNN is a vector that represents the semantic information of the artifact. The tracing network then compares the semantic vectors of two artifacts and outputs the probability that they are linked. 

Given the need for an initial training set of trace links, our approach cannot be used in an entirely green field domain. However, based on requests from our industrial collaborators, we envision the following  primary usage scenarios: (1) Train the tracing network on an initial set of manually constructed trace links for a project and then use it to automate the production of other links as the project proceeds; (2) Train the tracing network on the complete set of trace links for a project and then use it to find additional links that may have been missed during the manual link construction process; and finally (3) Train the tracing network on the trace links for one project, or for specific types of artifacts in one project, and then apply it to other projects and artifact types within the same domain. In this paper we focus on the first scenario. 

We evaluate our approach on a large industrial dataset taken from the domain of PTC systems to address two research questions: \vspace{2pt} \\
\noindent{\bf RQ1: }How should RNN be configured in order to generate the most accurate trace links?  \vspace{2pt} \\
\noindent{\bf RQ2: } Is RNN able to significantly improve trace link accuracy in comparison to standard baseline techniques? 


The remainder of the paper is structured as follows. We first introduce deep learning techniques related to the tracing network in Section \ref{sec:nnBackground}. The architecture of the tracing network is described in Section \ref{sec:traceModel}. Sections \ref{sec:experiment} and \ref{sec:result} describe the process used to configure the tracing network, our experimental design, and the results achieved.  Finally, in Sections \ref{sec:relatedWork} to \ref{sec:conclusion} we discuss related work, threats to validity, and conclusions.

\section{Deep Learning for Natural Language Processing}
\label{sec:nnBackground}
Many modern deep learning models and associated training methods originated from research in artificial neural networks (ANN). Inspired by advances in neuroscience, ANNs were designed to approximate complex functions of the human brain by connecting a large number of simple computational units in a multi-layered structure. Based on ANNs, \textit{Deep Learning} models feature more complex network connections in a larger number of layers. A benefit gained from  a more complex structure is the ability to represent data features with multiple levels of abstraction; this is usually preferable to more traditional machine learning techniques, in which human expertise is needed to select features of data for training. Back-propagation  \cite{rumelhart1986back_propogation} is widely recognized as an effective method for training deep neural networks; it indicates how the network should adapt its internal parameters to better compute the representation in each layer. Before presenting our approach, we describe fundamental concepts of deep learning techniques, especially as related to NLP tasks.  Furthermore, as our interest lies in comparatively evaluating different models for purposes of trace link creation, we describe these various techniques in some depth.

\subsection{Word Embedding}
\label{subsec:wordEmbedding}
Conventional NLP and information retrieval techniques treat unique words as atomic symbols and therefore do not take associations among words into account. To address this limitation, word embedding learns the representation of each word from a corpus as a continuous high dimensional vector, such that similar words are close together in the vector space. In addition, the embedded word vectors encode syntactic and semantic relationships between words as linear relationships between word vectors \cite{word2vec2013}. 
The use of learned word vectors is considered one of the primary reasons for the success of recent deep learning models for NLP tasks \cite{erhan2010pretraining}.

Skip-gram with negative sampling \cite{word2vec2013, DBLP:journals/corr/MikolovSCCD13} and GloVe \cite{pennington2014glove} are the most popular word embedding models due to the notable improvement they bring to word analogy tasks over more traditional approaches such as Latent Semantic Analysis \cite{word2vec2013, pennington2014glove}. Word embedding models are trained using unlabeled natural language text by utilizing co-occurrence statistics of words in the corpus. The Skip-gram model scans context windows across the entire training text to train prediction models \cite{word2vec2013}. Given the \emph{center} word in the window of size $T$, this model maximizes the probability that the targeted word appears around the center word, while minimizing the probability that a random word appears around the center word. The GloVe model uses matrix factorization; however we do not discuss it further because its performance is equivalent to the Skip-gram with negative sampling approach while it is less robust and utilizes more system resources \cite{levy2015improving}.
 

\subsection{Neural Network Structures}
Deep learning for NLP tasks are typically addressed using neural network techniques. \emph{Feedforward} networks, also referred to as multi-layer perceptrons (MLPs), represent a traditional neural network structure 
and lay the foundation for many other structures \cite{Haykin1994NNbook}. However, the number of parameters in a fully connected MLP can grow extremely large as the width and depth of the network increases. To address this limitation, researchers have proposed various neural network structures targeting different types of practical problems. For example, convolutional neural networks (CNNs) are especially well suited for image recognition and video analysis tasks \cite{lecun1989CNN}. For NLP tasks, Recurrent neural networks (RNNs) are widely used and are recognized as a good fit to the unique needs of NLP \cite{rumelhart1986back_propogation}. In particular, RNN and its variants have produced significant breakthroughs in many NLP tasks including language modeling \cite{mikolov2010recurrent}, machine translation \cite{DBLP:journals/corr/BahdanauCB14}, and semantic entailment \cite{tai2015tree_lstm}. In the following sections, we first introduce background about RNN and then discuss several RNN variants that were evaluated in this study.

\subsection{Standard  Recurrent  Neural  Networks  (RNN)}
\begin{figure}[h]
\centering
\includegraphics[width=0.46\textwidth]{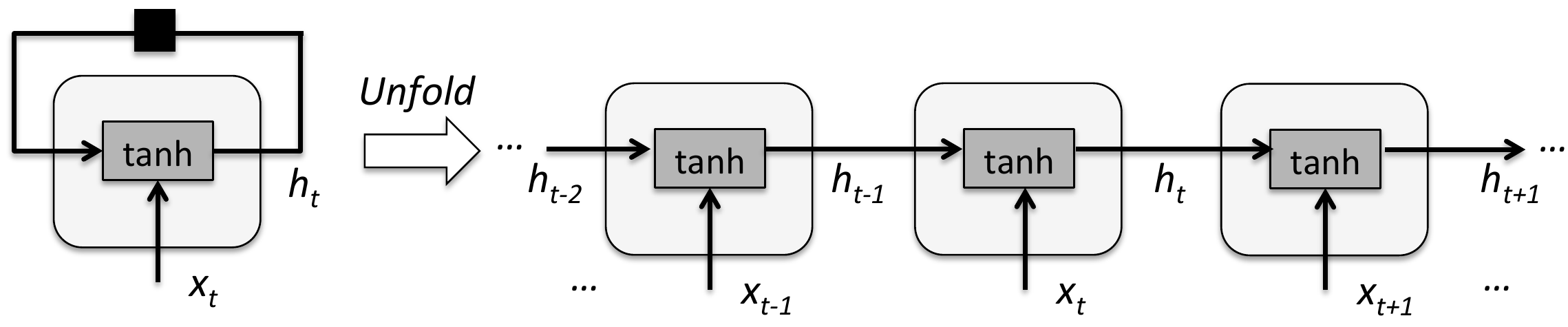}
\caption{Standard RNN model (left) and its unfolded architecture through time (right). The black square in the left figure indicates a one time step delay.}
\label{fig:rnn_model}
\vspace{-8pt}
\end{figure}

RNNs are particularly well suited for processing sequential data such as text and audio.  They connect computational units of the network in a directed cycle such that at each time step $t$, a unit in the RNN not only takes input of the current step (i.e., the next word embedding), but also the hidden state of the same unit from the previous time step $t-1$. This feedback mechanism simulates a ``memory'', so that a RNN's output is determined by both its current and prior inputs. Furthermore, because RNNs use the same unit (with the same parameters) across all time steps, they are able to process sequential data of arbitrary length.  This is illustrated in Figure \ref{fig:rnn_model}.
At a given time step $t$ with input vector $x_t$ and its previous hidden output vector $h_{t-1}$, a standard RNN unit calculates its output as \vspace{-2pt}
\begin{equation} \label{eq:rnn}
\begin{split}
h_t = & \ tanh(Wx_t +Uh_{t-1} + b)
\end{split}
\end{equation}

where $W$, $U$ and $b$ are the affine transformation parameters, and $tanh$ is the hyperbolic tangent function: $tanh(z) = (e^z -e^{-z})/ (e^z+e^{-z})$.

\vspace{-10pt}

A prominent drawback of the standard RNN model is that the network degrades when long dependencies exist in the sequence due to the phenomenon of exploding or vanishing gradients during back-propagation \cite{bengio1994learning}. This makes a standard RNN model difficult to train. The exploding gradients problem can be effectively addressed by scaling down the gradient when its norm is bigger than a preset value (i.e. \emph{Gradient Clipping}) \cite{bengio1994learning}. To address the vanishing gradients problem of the standard RNN model, researchers have proposed several variants with mechanics to preserve long-term dependencies; these variants included Long Short Term Memory (LSTM) and the Gated Recurrent Unit (GRU).

\label{sec:rnn}
\subsection{Long Short Term Memory (LSTM)}
LSTM networks include a memory cell vector in the recurrent unit to preserve long term dependencies\cite{hochreiter1997LSTM}. LSTM also introduces a gating mechanism to control when and how to read or write information to the memory cell. A gate in LSTM usually uses a sigmoid function $\sigma(z) = 1/(1+e^{-z})$ and controls information throughput using a point-wise multiplication operation $\odot$. Specifically, when the sigmoid function outputs $0$, the gate forbids any information from passing, while all information is allowed to pass when the sigmoid function output is $1$. Each LSTM unit contains an \emph{input} gate ($i_t$), a \emph{forget} gate ($f_t$), and an \emph{output} gate ($o_t$). The state of each gate is decided by $x_t$ and $h_{t-1}$ such that: 
\begin{equation} \label{eq:lstm_gate}
\begin{split}
i_t = & \ \sigma(W^ix_t + U^ih_{t-1} + b^i) \\
f_t = & \ \sigma(W^fx_t + U^fh_{t-1} + b^f)\\
o_t = & \ \sigma(W^ox_t + U^oh_{t-1} + b^o)\\
\end{split}
\end{equation}
 To update the information in the memory cell, a memory candidate vector $\tilde{c_t}$ is first calculated using the $tanh$ function. This memory candidate passes through the input gate, which controls how much each dimension in the candidate vector should be ``remembered''. At the same time, the \emph{forget} gate controls how much each dimension in the previous memory cell state $c_{t-1}$ should be retained. The actual memory cell state $c_t$ is then updated using the sum of these two parts.
\begin{equation} \label{eq:lstm_cell}
\begin{split}
\tilde{c_t} = & \ tanh(W^cx_t + U^ch_{t-1} + b^c)\\
c_t = & \ i_t \odot \tilde{c_t} + f_t \odot c_{t-1}\\
\end{split}
\end{equation}
Finally, the LSTM unit calculates its output $h_t$ with an output gate as follows: \vspace{-6pt}
\begin{equation} \label{eq:lstm_h}
\begin{split}
h_t = & \ o_t \odot tanh(c_t)\\
\end{split}
\end{equation}

Figure \ref{fig:lstm_vs_gru} (a) illustrates a typical LSTM unit. Using retained memory cell state and the gating mechanism, the LSTM unit ``remembers'' information until it is erased by the \emph{forget} gate; as such, LSTM handles long-term dependencies more effectively. LSTM has been repeatedly applied to solve semantic relatedness tasks and has achieved convincing performance \cite{tai2015tree_lstm, DBLP:journals/corr/RocktaschelGHKB15}. These advances motivated us to adopt LSTM for reasoning semantics in the tracing task.

\subsection{Gated Recurrent Unit (GRU)}
Finally, the recently proposed Gated Recurrent Unit (GRU) model also uses a gating mechanism to control the information flow within a unit; but it has a simplified unit structure and does not have a dedicated memory cell vector \cite{cho2014gru}. It contains only a \emph{reset} gate $r_t$ and an \emph{update} gate $u_t$:\vspace{-2pt}
\begin{equation} \label{eq:gru_gate}
\begin{split}
r_t = & \ \sigma(W^rx_t + U^rh_{t-1} + b^r)\\
u_t = & \ \sigma(W^ux_t + U^uh_{t-1} + b^u)\\
\end{split}
\end{equation}

In GRU networks, the previous hidden output $h_{t-1}$ goes through the \emph{reset} gate $r_t$ and is sent back to the unit. An output candidate $\tilde{h_t}$ is then calculated using the gated $h_{t-1}$ and the unit's current input $x_t$ as follows:
\begin{equation} \label{eq:gru_h_cand}
\begin{split}
\tilde{h_t} = & \ \tanh(W^hx_t + U^h(r_t \odot h_{t-1}) + b^h) \\
\end{split}
\end{equation}
The actual output of a unit $h_t$ is a linear interpolation between the previous output $h_{t-1}$ and the candidate output $\tilde{h_t}$, controlled by the \emph{update} gate $u_t$. As such, the \emph{update} gate balances how much of the current output is updated using $\tilde{h_t}$ and $h_{t-1}$.
\begin{equation} \label{eq:gru_h}
\begin{split}
h_t = & \ (1-u_t) \odot h_{t-1} + u_t \odot \tilde{h_t} \\
\end{split}
\end{equation}

\begin{figure}[t!]
\centering
\begin{minipage}{.28\textwidth}
  \centering
  \includegraphics[width=.95\textwidth]{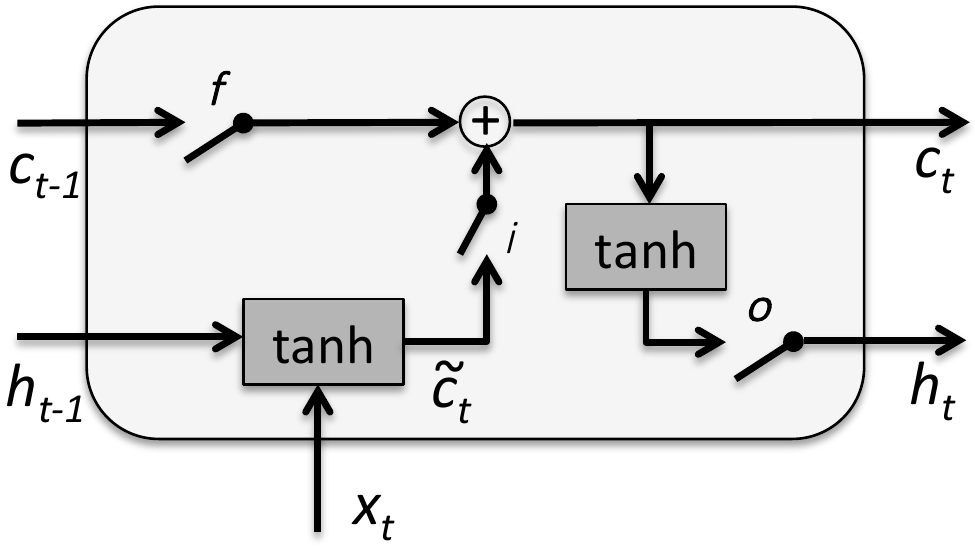}
  \subcaption{LSTM}
  \label{fig:lstm_model}
\end{minipage}%
\begin{minipage}{.22\textwidth}
  \centering
  \includegraphics[width=.95\linewidth]{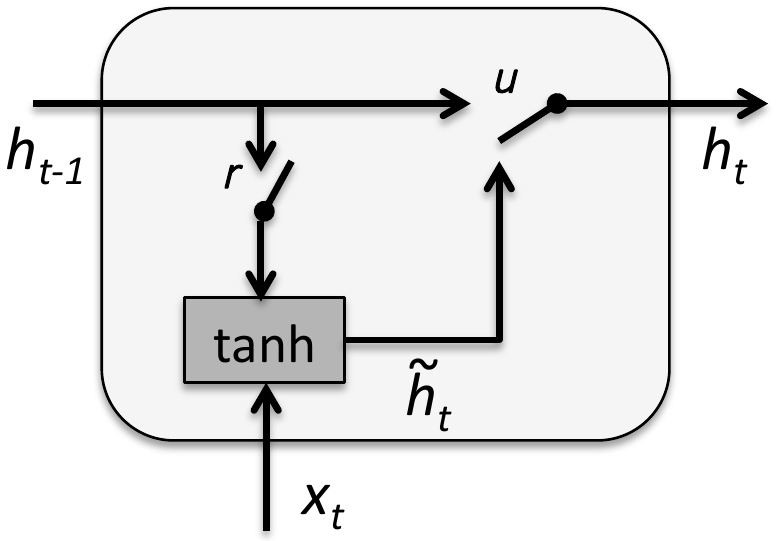}
  \subcaption{GRU}
  \label{fig:gru_model}
\end{minipage}
\caption{Comparison between single units from LSTM and GRU networks. In (\ref{fig:lstm_model}), $i$, $f$ and $o$ are the input, forget and output gates respectively; $c$ is the memory cell vector. In (\ref{fig:gru_model}), $r$ is the reset gate and $u$ is the update gate  \cite{chung2014eval_gru}.}
\label{fig:lstm_vs_gru}
\vspace{-12pt}
\end{figure}

Consequently, a GRU unit embeds long-term information directly into the hidden output vectors. Figure \ref{fig:lstm_vs_gru} compares the unit structures of LSTM and GRU networks. Despite having a simpler structure, GRU has achieved competitive results with LSTM for many NLP tasks \cite{DBLP:journals/corr/BahdanauCB14, chung2014eval_gru}, and as such no decisive conclusion has been drawn about which model is better. In this work, we compare the performance of LSTM and GRU in order to identify the most suitable model for addressing the tracing problem. 

\subsection{Other RNN Variables}
In addition to modifying the structure within a single RNN unit, the structure of the overall RNN network can be varied. For instance, multi-layered RNNs \cite{PascanuGCB2013deeprnn} stack more than one RNN unit at each time step \cite{schmidhuber1992learning} with the aim of extracting more abstract features from the input sequence. In contrast, bi-directional RNNs \cite{schuster1997birnn} process sequential data in both forward and backward directions at the same time; this enables the output to be influenced by both past and future data in the sequence. In this study, we also explored the use of two-layered RNNs and bidirectional RNNs for generating trace links. 

\section{The Tracing Network}
\label{sec:traceModel}
The tracing process comprises several steps: first, the human analyst initiates a trace for a source artifact; second, the similarity between the source artifact and each of the potentially linked target artifacts is computed; third, a list of ranked candidate links are returned; and finally the human evaluates the links and accepts the ones deemed to be correct.  The process is repeated for all source artifacts \cite{DBLP:journals/tse/HayesDS06}. Out study investigates the effectiveness of various deep learning models and methods for calculating similarities between source and target artifact pairs, with the goal of generating accurate trace links. This is essentially a textual comparison task in which the tracing network needs to leverage domain knowledge to understand the semantics of two individual artifacts and then to evaluate their semantic relatedness for tracing purposes. Valid associations need to be established between related artifacts even when no common words are present. Based on our initial analysis of the strengths and weaknesses of current techniques, we decided to adopt word embeddings and RNN techniques to achieve this goal. Therefore, we first need to learn word embeddings from a domain corpus in order to effectively encode word relations, and then utilize such word embeddings in the tracing network structure to extract and compare their semantics. In this section, we describe our approach for designing and training such a tracing network.

\vspace{-6pt}
\begin{figure}[h]
\centering
\includegraphics[width=0.48\textwidth]{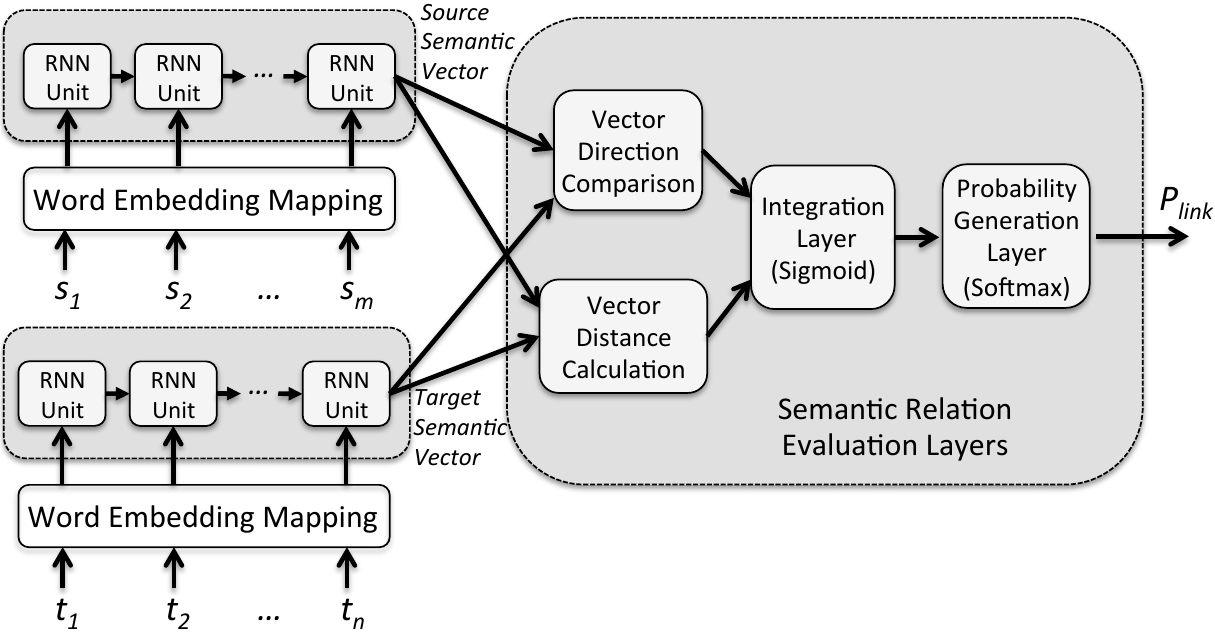}
\caption{The architecture of the tracing network. The software artifacts are first mapped into sequences of embedded word vectors and go through RNN layers to generate the semantic vectors, which are then fed into the Semantic Relation Evaluation layers to predict the probability that they are linked.}
\label{fig:trace_model}
\vspace{-8pt}
\end{figure}


\subsection{Network Architecture}
The design of the neural network architecture is shown in Figure \ref{fig:trace_model}. Given the textual content of a source artifact $A_s$ and a target artifact $A_t$, each word in $A_s$ and $A_t$ is first mapped onto its vector representation through the \textit{Word Embedding} layer. Such mappings are trained from the domain corpus using the Skip-gram model introduced in Section \ref{subsec:wordEmbedding}. The vectors of words in source artifact $s_1$, $s_2$, \dots, $s_m$ are then sent to the \textit{RNN} layers sequentially and output as a single vector $v_s$ representing its semantic information. In the case of the bidirectional-RNN, the word vectors are also sent in reverse order as $s_m$, $s_{m-1}$, \dots, $s_1$. The target semantic vector $v_t$ is generated in the same way using RNN layers. Finally, these two vectors are compared in the \textit{Semantic Relation Evaluation} layers.

The \textit{Semantic Relation Evaluation} layers in our tracing network adopt the structure proposed by Tai et al. \cite{tai2015tree_lstm}, targeted to perform semantic entailment classification tasks for sentence pairs. The overall calculation of this part of the network can be represented as:
\begin{equation} \label{eq:semantic_network}
\begin{split}
r_{pmul} = & \ v_s \odot v_t \\
r_{sub} = & \ |v_s - v_t| \\
r = & \ \sigma(W^{r}r_{pmul} + U^{r}r_{sub} + b^r) \\
p_{tracelink} = & \ softmax(W^pr +b^p) \\
\end{split}
\end{equation}
where
\vspace{-6pt}
\begin{equation} \label{eq:softmax}
softmax(z)_j = e^{z_j}/\sum_{k=1}^Ke^{z_k}, for j = 1,\dots,K \end{equation}
Here, $\odot$ is the point-wise multiplication operator used to compare the \emph{direction} of source and target vectors on each dimension. The absolute vector subtraction result, $r_{sub}$, represents the \emph{distance} between the two vectors in each dimension. The network then uses a hidden \emph{sigmoid layer} to integrate $r_{pmul}$ and $r_{sub}$ and output a single vector to represent their semantic similarity. Finally, the output \emph{softmax layer} uses the result to produce the probability that a valid trace link exists; the result of a softmax function is a K-dimensional vector of real values in the range $(0, 1)$ that add up to 1 (K=2 in this case).

A concrete tracing network is built upon this architecture and is further configured by a set of network settings. Those settings specify the type of RNN unit (i.e. GRU or LSTM), the number of hidden dimensions in RNN units and the Semantic Relation Evaluation layers, and other RNN variables such as the number of RNN layers and whether to use bidirectional RNN. To address our first research question (RQ1) we explored several different configurations. We describe how we optimized network settings in Section \ref{subsec:networkOptimization}. The tracing network is implemented on the Torch framework (http://torch.ch), and the source code is available at https://github.com/jin-guo/TraceNN.

\subsection{Training the Tracing Network}
A powerful network is only useful when it can be properly trained using existing data and when it is generalizable to unseen data. To train the tracing network, we use the regularized negative log likelihood as our objective loss function to be minimized. This objective function is commonly used in categorical prediction models \cite{Goodfellow-et-al-2016-Book} and can be written as:
\begin{equation} \label{eq:loss}
\begin{split}
J(\theta) = -\frac{1}{N}\sum_{i=1}^NlogP(Y=y^i|x^i,\theta) + \frac{\lambda}{2}\|\theta\|_2^2
\end{split}
\end{equation}
where $\theta$ indicates the network parameters that need to be trained, $N$ is the total number of examples in the training data,  $x^i$ is the input of the \textit{ith} training example,  $y^i$ is the actual category of that example (i.e. link or non-link); as a result, $P(Y=y^i|x^i,\theta)$ represents the network's prediction on the correct category given the current input and parameters. The second part of the loss function represents a $L^2$ parameter regularization that prevents overfitting, where $\|\theta\|_2$ is the Euclidean norm of $\theta$, and $\lambda$ controls the strength of the regularization.

Based on this loss function, we used a stochastic gradient descent method \cite{tieleman2012rmsprop} to update the network parameters. According to this method, a typical training process is comprised of a number of epochs. Each \textit{epoch} iterates through all of the training data one time to train the network; as such, the overall training process uses all the training data several times until the objective loss is sufficiently small or fails to decrease further. In each epoch, the training data is further randomly divided into a number of ``mini batches;'' each contains one or more training datapoints. After each batch is processed, a gradient of parameters is calculated based on the loss function. The network then updates its parameters based on this gradient and a ``learning rate'' that specifies how fast the parameters move along the gradient direction. During training, we adopted an adaptive learning rate procedure \cite{tieleman2012rmsprop} to adjust the learning rate based on the current training performance. To help the network converge, we also decreased the learning rate after each epoch until epoch $\tau$, such that the learning rate at epoch $k$ is determined by:
\begin{equation} \label{eq:learningRate}
\epsilon_k = (1-\alpha)\epsilon_0 + \alpha\epsilon_\tau
\end{equation}
where $\epsilon_0$ is initial learning rate, $\alpha = k/\tau$. In our experiment, $\epsilon _\tau$ is set to $\epsilon_0/100$, and  $\tau$ set to 500.

Based on these general methods, a tracing network training process is further determined by a set of predefined \emph{hyper-parameters} that help steer the learning behavior. Common hyper-parameters include the initial learning rate, gradient clip value, regulation strength ($\lambda$), the number of datapoints in a mini batch (i.e. mini batch size), and the number of epochs included in the training process. The techniques for selecting hyper-parameters are described in Section \ref{subsec:networkOptimization}.

\section{Experiment Setup}
\label{sec:experiment}
In this section, we explain the methods used to (1) prepare data, (2) systematically tune the configuration (i.e. network settings and hyper-parameters) of the tracing network, and (3) compare the performance of the best configuration against other popular trace evaluation methods.

\subsection{Data Preparation}
\label{subsec:dataPreparation}
To train the word embeddings, we used a corpus from the PTC domain that is comprised of 52.7MB of clean text extracted from related domain documents and software artifacts. The original corpus of domain documents was collected from the Internet by our industrial collaborators as part of their initial domain analysis process.  We also added the latest Wikipedia dump containing about 19.92GB of clean text to the corpus and used it for one variant of the word embedding configuration.  All documents were preprocessed by transforming characters to lower-case and removing all non-alpha-numeric characters except for underscores and hyphens.  

To train and evaluate other parts of the tracing network, we used PTC project data provided by our industrial collaborators. The dataset contains 1,651 Software Subsystem Requirements (SSRS) as source artifacts and 466 Software Subsystem Design Descriptions (SSDD) as target artifacts. Each source artifact contained an average of 33 tokens and described a functional requirement of the Back Office Server (BOS) subsystem. Each target artifact contained an average of 99 tokens and specified design details. There were 1,387 trace links between SSRS and SSDD artifacts, all of which were constructed and validated by our industrial collaborators. This dataset is considerably larger than those used in most previous studies on requirements traceability \cite{DBLP:conf/icse/AsuncionAT10, lucia2007recovering, DBLP:journals/tse/HayesDS06} and the task of creating links across such a large dataset represents a challenging industrial-strength tracing problem. 
We randomly selected 45\% of the 769,366 artifact pairs from the PTC project dataset (i.e. 1,651 $\times$ 466) for inclusion in a training set, 10\% for a development set, and 45\% as a testing set. Given a fixed tracing network configuration, the training set was used to update the network parameters (i.e. the weight and bias for affine transformation in each layer) in order to minimize the objective loss function. The development set was used to select the best general model during an initial training process to ensure that the model was not overtrained. The test data was set aside and only used for evaluating the performance of the final network model. 

Software project data exhibits special characteristics that impact the training of a neural network \cite{DBLP:conf/msr/0004RCRHV16}. In particular, the number of actual trace links is usually very small for a given set of source and target artifacts compared to the total number of artifact pairs. In our dataset, among all 769,366 artifact pairs, only 0.18\% are valid links. Training a neural network using such an unbalanced dataset is very challenging. A common and relatively simple approach for handling unbalanced datasets is to weight the minority class (i.e. the linked artifacts) higher than the majority class (i.e. the non-linked ones). However, in the gradient descent method, a larger loss weighting could improperly amplify the gradient update for the minority class making the training unstable and causing failure to converge. Another common way to handle unbalanced data is to downsample the majority case in order to produce a fixed and balanced training set. Based on initial experimentation we found that this approach did not yield good results because the examples of non-links used for training the network tended to randomly exclude artifact pairs that lay at the frontier at which links and non-links are differentiated. Furthermore, based on initial experimentation, we also ruled out the upsampling method because this considerably increased the size of the training set, excessively prolonging the training time.

Based on further experimentation we adopted a strategy that dynamically constructed balanced training sets using sub-datasets. In each epoch, a balanced training set was constructed by including all valid links from the original training set as well as a randomly selected equal number of non-links form the training set. The selection of non-links was updated at the start of each epoch. This approach ensured that over time the sampled non-links used for training were representative and preserved an equal contribution of links and non-links during each epoch. Our initial experimental results showed this technique to be effective for training our tracing network.

\subsection{Model Selection and Hyper-Parameters Optimization}
\label{subsec:networkOptimization}
Finding suitable network settings and a good set of hyper-parameters is crucial to the success of applying deep learning methods to practical problems \cite{pinto2009high}. However, given the running time required for training, the search space of all the possible combinations of different configurations was too large to provide full coverage. We therefore first identified several configurations that were expected to produce good performance. This was accomplished by manually observing how training loss changed during early epochs and following heuristics suggested in \cite{bengio2012practical}. We then created a network configuration search space centered around these manually identified configurations; our search space is summarized in table \ref{tab:hyperparam}. We conducted a \emph{grid search} and trained all the combinations of each configuration in Table \ref{tab:hyperparam} using the training set and then compared their performance on the development set to find the best configuration. We describe our search space below.

\begin{table}
  \caption{Tracing Network Configuration Search Space}
  \label{tab:hyperparam}
    \begin{center}
        \begin{tabular}{ | l | p{5cm} |}
        \hline
        Word Embedding Source &  PTC docs -- 50 dim, \\
        & PTC docs + Wikipedia dump -- 300 dim  \\ \hline \hline
        RNN Unit Type & GRU, LSTM, BI-GRU, BI-LSTM,\\ 
        &  (AveVect as baseline) \\ \hline
        RNN Layer & 1, 2  \\ \hline 
        Hidden Dimension & RNN30 + Intg10, RNN60 + Intg20 \\ \hline \hline
        Init Learning Rate ($lr$) & 1e-03, 1e-02, 1e-01 \\ \hline
        Gradient Clip Value ($gc$) & 10, 100  \\ \hline
        Regularization Strength $\lambda$ & 1e-04, 1e-03  \\ \hline \hline
        Mini Batch Size & 1  \\ \hline
        Epoch & 60  \\ \hline
        \end{tabular}
    \end{center}
    \vspace{-16pt}
\end{table}

For learning the word embeddings, we used the Skip-gram model provided by the Word2vec tool \cite{word2vec2013}. We trained the word vectors with two settings: 50-dimension vectors using the PTC corpus only and 300-dimension using both PTC and Wikipedia dump. The number of dimensions are set differently because the PTC corpus (38,771 tokens) contains considerably less tokens than the PTC + Wikipedia dump (8,025,288 tokens). While a smaller vector dimension would result in faster training, a larger dimension is needed to effectively represent the semantics of all tokens in the latter corpus.

To compare which variation of RNN best suits the tracing network, we evaluated GRU, LSTM, bi-directional GRU (BI-GRU), bi-directional LSTM (BI-LSTM) with both 1 and 2 layers introduced in Section \ref{sec:rnn}. The hidden dimensions in each RNN unit were set to either 30 or 60, while the hidden dimensions for the Integration layer were set to 10 or 20 correspondingly. As a baseline method, we also replaced the RNN layers with a bag-of-word method in which the semantic vector of an artifact is simply set to be the average of all word vectors contained in the artifact (``AveVect'' in Table \ref{tab:hyperparam}). We also summarize the search space for other hyper-parameters of the tracing network in Table \ref{tab:hyperparam}. 


\subsection{Comparison of Tracing Methods}
In practical requirements tracing settings, a tracing method returns a list of candidate links between a source artifact, serving the role of user query, and a set of target artifacts. An effective algorithm would return all valid links close to the top of the list.  The effectiveness of a tracing algorithm is therefore often measured using Mean Average Precision (MAP).  To calculate MAP, we first calculate the Average Precision (AP) for each individual query as:
\begin{equation} \label{eq:AvgPrec}
AP = \frac{\sum _{i=1}^{|Retrieved|~} (Precision(i) \times relevant(i))}{|RelevantLinks|}
\end{equation}
where $|RetrievedLinks|$ is the number of retrieved links, $i$ is the rank in the sequence of retrieved candidates links, $relevant(i)$ is a binary function assigned 1 if the link is valid and 0 otherwise, and $Precision(i)$ is the precision computed after truncating the list immediately below $i$. Then, Mean Average Precision (MAP) is computed as the mean AP across all queries. In typical information retrieval settings, MAP is computed for the top N returned links; however, for traceability purposes we compute it when returning {\bf all} valid links as specified in the trace matrix.  This means that our version of MAP is computed for recall of 100\%. 

We computed MAP using the test dataset only, and compared the performance of our tracing network with other popular tracing methods, i.e. Vector Space Model (VSM) and Latent Semantic Indexing (LSI). To make a fair comparison, we also optimized the configurations for the VSM and the LSI methods using a Genetic Algorithm to search through an extensive configuration space of preprocessors and parameters \cite{DBLP:conf/sigsoft/LoharAZC13}. Finally, we configured VSM to use a local Inverse Document Frequency (IDF) weighting scheme when calculating the cosine similarity \cite{DBLP:journals/tse/HayesDS06}. LSI was reduced to 75\% dimensions. For both VSM and LSI we preprocessed the text to remove non alpha-numeric characters, remove stop words, and to stem each word using Porter's stemming algorithm.  

We also evaluated the results by plotting a precision vs. recall curve. The graph depicts \emph{recall} and \emph{precision} scores at different similarity or probability values. The Precision-Recall Curve thus shows  trade-offs between precision and recall and provides insights into where each method performs best -- for example, whether a technique improves precision at higher or lower levels of recall \cite{buckland1994relationship}.  A curve that is farther away from the origin indicates better performance.

\section{Results and Discussion}
\label{sec:result}
In this section, we report (1) the best configurations found in our network configuration search space, (2) the performance of the tracing network with the best configuration compared against VSM and LSI, and (3) the performance of the tracing network when trained with a larger training set of data. 

\subsection{What is the best configuration for the tracing network?}
\label{subsec:exp1}

This experiment aims to address the first research question (RQ1). When optimizing the network configuration of the tracing network, we first selected the best configuration for each RNN unit type; Table \ref{tab:BestHyperparamPerModel} summarizes these results. We found that the best configurations for all four RNN unit types were very similar: one layer RNN model with 30 hidden dimensions and an Integration layer of 10 hidden dimensions, learning rate of 1e-02, gradient clip value of 10, and $\lambda$ of 1e-04. Performance varies for different RNN unit types.

\begin{table}[h]
  \caption{Best Configuration for Each RNN Unit Type}
  \label{tab:BestHyperparamPerModel}
    \begin{center}
        \begin{tabular}{p{1.2cm}p{0.6cm}p{1cm}llllll}
            \toprule
            \thead{RNN\\Unit} & \thead{Dev.\\Loss} & \thead{Word\\Emb.} & \thead{$L$} & \thead{$D_r$} & \thead{$D_s$} & \thead{$lr$} & \thead{$gc$} & \thead{$\lambda$}\\
            \midrule
            BI-GRU  & .1045     & PTC       &1    & 30   & 10 & .01  & 10      & .0001 \\
            GRU     & .1301     & PTC       &1    & 30   & 10 & .01  & 10      & .0001 \\
            BI-LSTM & .1434     & PTC       &1    & 30   & 10 & .01  & 100     & .0001 \\
            LSTM    & .2041     & PTC+Wiki  &1    & 30   & 10 & .01  & 10      & .0001 \\
            \bottomrule
        \multicolumn{9}{p{8.5cm}}{$L$ -- number of layers in the RNN model, $D_r$ -- hidden dimension in RNN unit, $D_s$ -- hidden dimension in Integration layer, $lr$ -- initial learning rate, $gc$ -- gradient clipping value, $\lambda$ -- regularization strength}
        \end{tabular}
    \end{center}
    \vspace{-12pt}
\end{table}

\begin{figure}[h]
\centering
\includegraphics[width=0.49\textwidth]{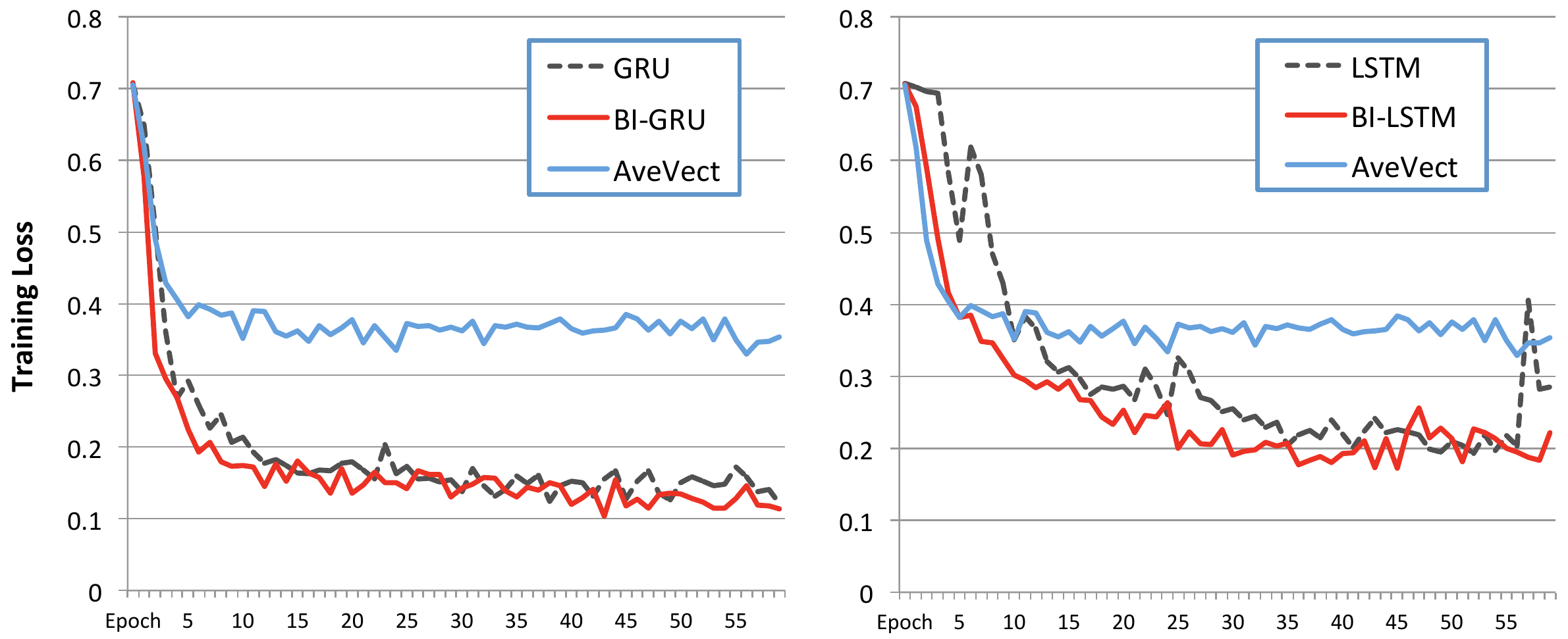}
\caption{Comparison of learning curves for RNN variants using their best configurations. GRU and BI-GRU (left) converged faster and achieved smaller loss than LSTM and BI-LSTM (right). They all outperformed the baseline.}
\label{fig:learningCurve}
\vspace{-4pt}
\end{figure}

Figure \ref{fig:learningCurve} illustrates the learning curves on the training dataset of the four configurations. All four RNN unit types outperformed the Average Vector method. This supports our hypothesis that word order plays an important role when comparing the semantics of sentences. Both GRU and BI-GRU achieved faster convergence and more desirable (smaller) loss than LSTM and BI-LSTM. Although quite similar, the bidirectional models performed slightly better than the unidirectional models for both GRU and LSTM on the training set; the bidirectional models also achieved better results on the development dataset compared to their unidirectional counterparts. As a result, the overall best performance was achieved using \textbf{BI-GRU} configured as shown in Table \ref{tab:BestHyperparamPerModel}.

We also found that in three of the four best configurations, the word embedding vectors were trained using the PTC corpus alone. We speculate that one reason the PTC-trained word vectors performed better than the PTC+Wiki-trained vectors is due to differences in content of the two corpora. The PTC+Wiki corpus contains significantly more words that are used in diverse contexts because the majority of articles in the Wiki corpus are not related to the PTC domain. In the case of words that appear commonly in Wiki articles but convey specific meanings in the PTC domain (e.g. message, administrator, field, etc.), their context in more general articles is likely to negatively affect the reasoning task on domain specific semantics when there is insufficient training data to disambiguate their usage. We also tested the tracing network performance using 300-dimension word embedding trained by the PTC corpus. The result is similar to the best configuration found in Table \ref{tab:BestHyperparamPerModel}. These findings suggest that using only the domain corpus to train word vectors with a reasonable size is more computationally economical and can yield better results.

\subsection{Does the tracing network outperform leading trace retrieval algorithms?}
\label{subsec:exp2}
We now evaluate whether the best configuration of the tracing network (i.e. BI-GRU with configuration shown in Table \ref{tab:BestHyperparamPerModel}) outperforms leading trace retrieval algorithms. Links were therefore generated for each  source and target artifact pair in the test set (45\% total data) and for each source artifact ranked by descending probability scores. Average Precision ($AP$) was calculated using Equation \ref{eq:AvgPrec}. We then compared the APs for our tracing network against those generated using the best performing VSM and LSI configurations. Our tracing network was able to achieve a MAP of .598; this value is 41\% higher than that achieved using VSM (\emph{MAP} = .423) and 32\% higher than LSI (\emph{MAP} = .451). We conducted a Friedman test and found a statistically significant difference among the AP values associated with the three methods ($\mathcal{X}^2(2)=89.40$, $p<.001$). We then conducted three pairwise Wilcoxon signed ranks tests with Bonferroni p-value adjustments. Results indicated that the APs associated with our tracing network were significantly higher (\emph{M} = .598, \emph{SD} = .370) than those achieved using VSM (\emph{M} = .423, \emph{SD} = .391; $p<.001$) and LSI (\emph{M} = .451, \emph{SD} = .400; $p<.001$); in contrast, there was no significant difference when comparing APs for VSM versus those for LSI.

\begin{figure}[t]
\centering
\includegraphics[width=0.45\textwidth]{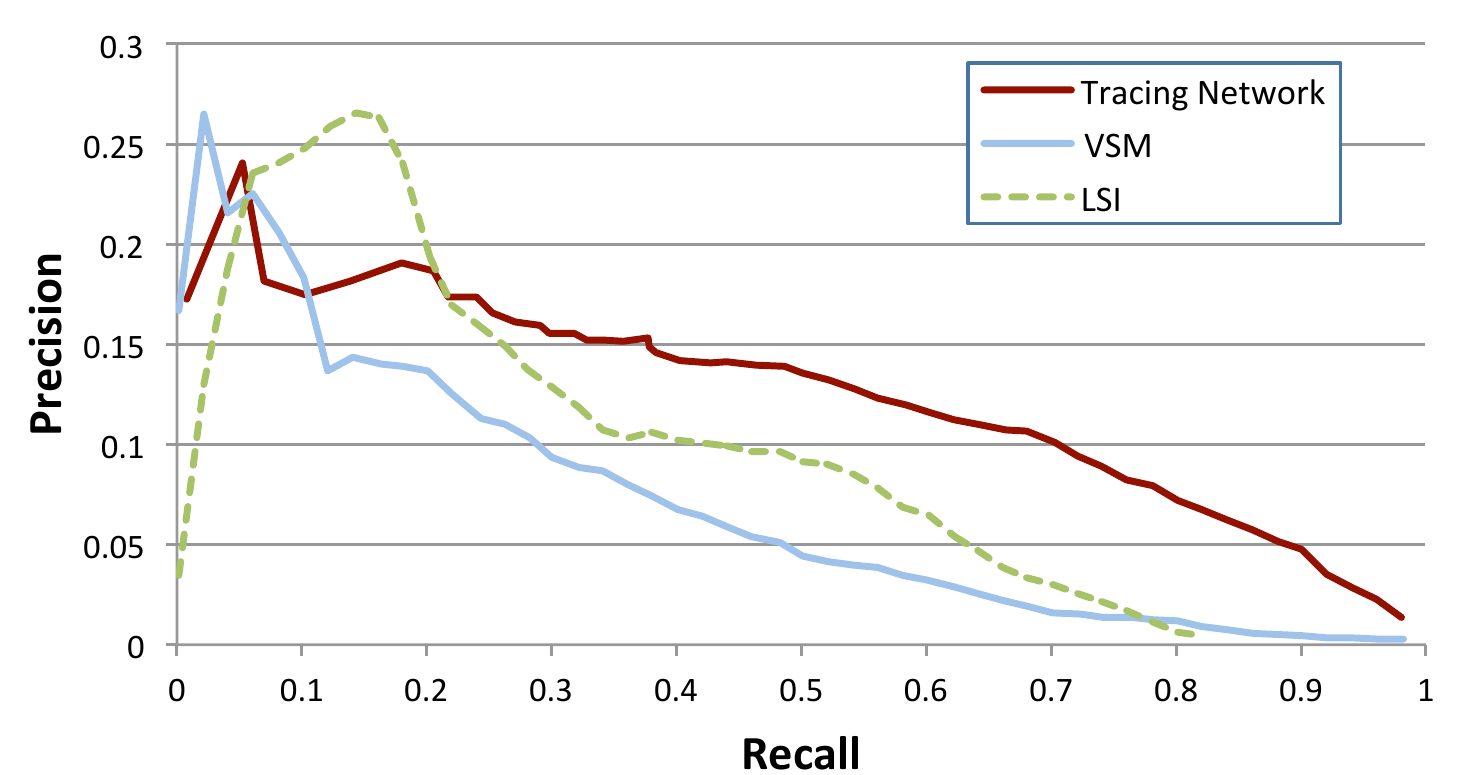}
\caption{Precision-Recall Curve on test set -- 45\% total data}
\label{fig:prCurve_45_45_10}
\end{figure}

When comparing the Precision-Recall Curves for the three methods (Figure \ref{fig:prCurve_45_45_10}), we observed that the tracing network outperformed VSM and LSI at higher levels of recall. Given the goal to achieve close to 100\% recall when performing tracing tasks, this is an important achievement. Precision improved notably when recall was above 0.2. This improvement can be attributed to the fact that the tracing network is able to extract semantic information from artifacts and to reason over associations between them.

\begin{figure}[t]
    \centering
    \begin{subfigure}[b]{0.30\textwidth}
        \includegraphics[width=\textwidth]{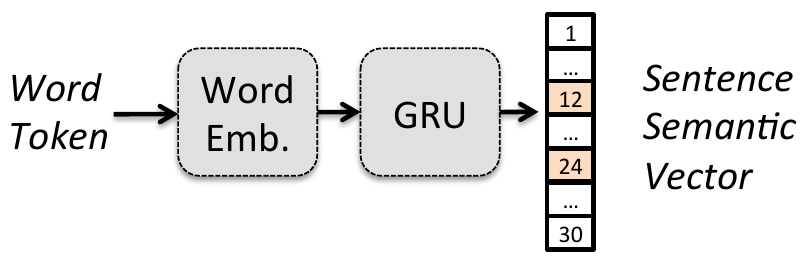}
        \caption{GRU output at one time step}
        \label{fig:snippet}
    \end{subfigure}
    
    \vspace{0.3cm}

    \begin{subfigure}[b]{0.49\textwidth}
        \includegraphics[width=\textwidth]{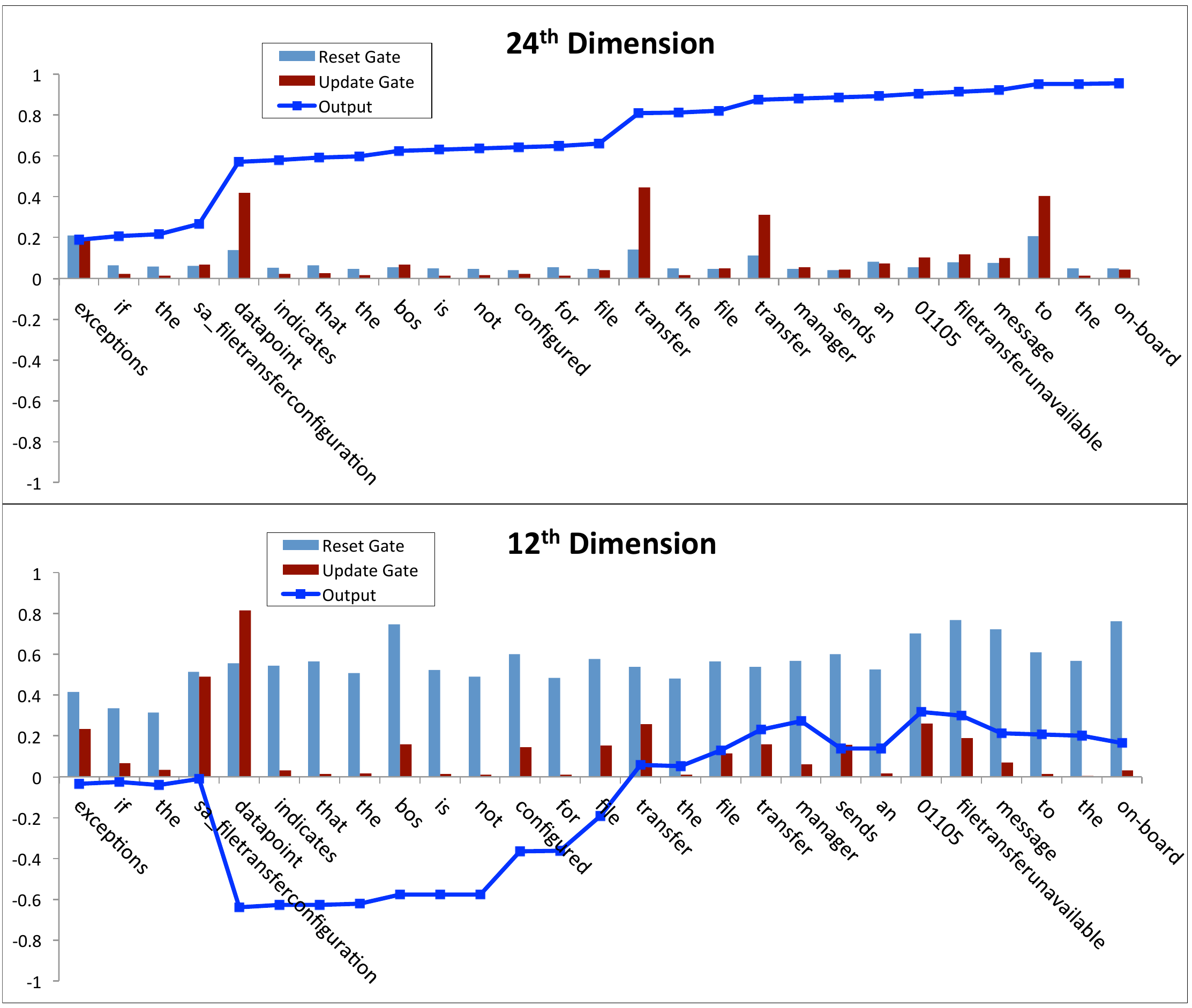}
        \caption{GRU gate behavior}
        \label{fig:gateBehavior}
    \end{subfigure}

    \caption{The reset gate and update gate behavior in GRU and their corresponding output on the 24th and 12th dimension of sentence semantic vector. The x-axis indicates the sequential input of words while the y-axis is the value of reset gate, update gate and output after taking each word.}\label{fig:gateOutput}
    \vspace{-16pt}
\end{figure}

While it is almost impossible to completely decipher how GRU extracts a semantic vector from natural language \cite{DBLP:journals/corr/KarpathyJL15}, observing gate behavior when GRU processes a sentence can provide some insights into how GRU performs this task so well.
As an example, we examine the gate behavior when our best performing GRU processes the following artifact text: \textit{``Exceptions: If the SA\_FileTransferConfiguration datapoint indicates that the BOS is not configured for file transfer, the File Transfer Manager sends an 01105\_FileTransferUnavailable message to the On\-board.''}. The gate behavior varies among dimensions in the sentence semantic vector. In Figure \ref{fig:gateOutput}, we show the levels of the reset and the update gates and the change of the output values on two of the 30 dimensions after GRU takes each word from this artifact. 
Unlike LSTM, the GRU does not have an explicit memory cell; however, the unit output itself can be considered as a \emph{persisting memory}, whereas the output candidate in Equation \ref{eq:gru_h_cand} acts as a \emph{temporary memory}. Recall that the \textbf{reset gate} controls how much previous output adds to the current input when determining the output candidate (i.e. temporary memory); in contrast, the \textbf{update gate} balances the contributions of this temporary memory and the previous output  (i.e. persisting memory) to the actual current output. From Figure \ref{fig:gateBehavior}, we observe that for the 24th dimension, the reset gate is constantly small. This means that the temporary memory is mostly decided by the current input word. The update gate for this dimension was also small for most of the input words until the words \textit{datapoint},  \textit{transfer} and \textit{to} came. Those spikes indicate moments when the temporary memory was integrated into the persisting memory. As such, we can speculate that this semantic vector dimension functions to accumulate  information from specific keywords across the entire sentence. Conversely, for the 12th dimension, the reset gate was constantly high, indicating that the information stored in the temporary memory was based on both previous output and current input. Therefore, the actual output is more sensitive to local context rather than a single keyword. This is confirmed by the fluctuating shape of the actual output shown in the figure. For example, the output value remained in the same range until the topic was changed from ``datapoint indication'' to ``message transfer''. We believe that this versatile behavior of the gating mechanism might enable GRU to encode complex semantic information from a sentence into a vector to support trace link generation and other challenging NLP tasks.  

From Figure \ref{fig:prCurve_45_45_10}, we also notice that the tracing network hits a glass ceiling for improving precision above 0.27. We consider this to be caused by its inability to rule out some false positive links that contain valid associations. For example, the tracing network assigns a 97.27\% probability of a valid link between artifact \textit{``The BOS administrative toolset shall allow an authorized administrator to view internal errors generated by the BOS''} and the artifact \textit{``The MessagesEvents panel provides the functionality to view message and event logs.  The panel provides searching and filtering capabilities where the user can search by a number of parameters depending on the type of data the user wants to view''}. There are direct associations between these artifacts: MessagesEvents panel is part of the BOS administrative toolset for viewing message and event log, administrator is a system user and internal errors generated by the BOS is an event. But this is not a valid link because MessagesEvents panel only displays messages and events related to external Railroad Systems rather than to internal events. It is likely that the tracing network fails to exclude this link because it has not been exposed to sufficient similar negative examples in the training data. As we described in Section \ref{subsec:dataPreparation}, every positive example is used while negative examples (i.e. non-links) are randomly selected during each epoch. We plan to explore more adequate methods for handling unbalanced data problems caused by characteristics of the tracing data in our future work.

\subsection{How does the tracing network react to more training data?}
\label{subsec:exp3}
The number of trace links tends to increase as a software project evolves. To explore the potential impact of folding them into the training set, we increased the training dataset to 80\%. We randomly selected part of the test data and moved it into the training set to reach 80\%, while retaining the remaining data in the testing set.  Using the same configuration as described in Section \ref{subsec:networkOptimization}, we then retrained the tracing network. Because the size of the test set decreased to 10\%, we could not make direct comparisons to our previous results.  Instead we used both of the trained tracing networks (i.e. trained with 45\% and 80\% of the data respectively) to generate trace links against the same small test set (sized at 10\%). To reduce the effect of random data selection, we repeated this process five times and report the average results.

\begin{figure}[t!]
\centering
\includegraphics[width=0.46\textwidth]{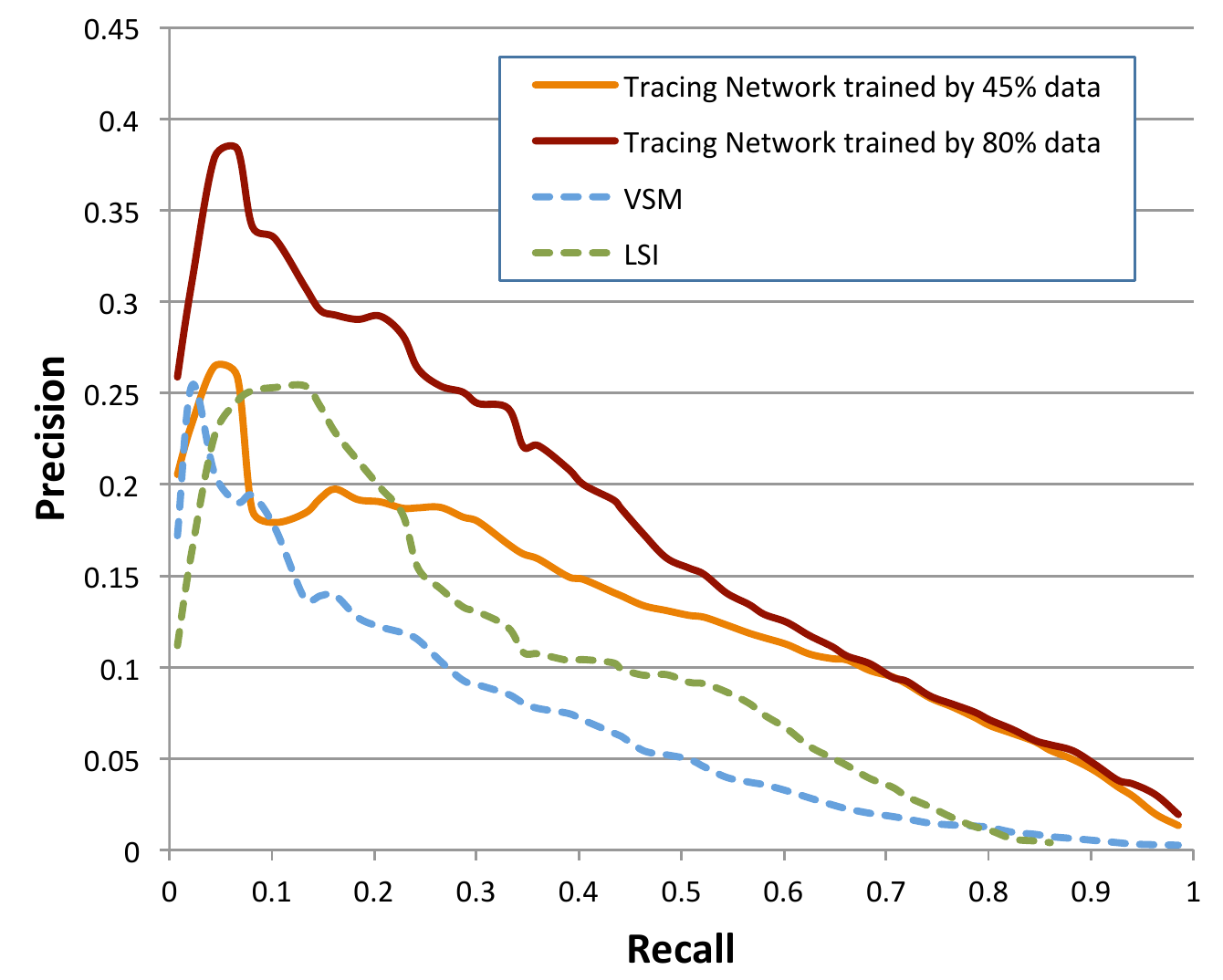}
\caption{Precision-Recall Curve on test set -- 10\% total data.}
\label{fig:PRCurve_before_after}
\vspace{-16pt}
\end{figure}

With a larger training set, the \emph{MAP} was .834 compared to .803 for the smaller training set. Results are  depicted in the Precision-Recall Curve in Figure \ref{fig:PRCurve_before_after}. With increased training data, the network can better differentiate links and non-links, and therefore improve both precision and recall. Improvements were observed especially at low levels of recall. 

The performance of the tracing network trained using 80\% of data was again compared against VSM and LSI for this larger training set. A Friedman test identified a statistically significant difference among the APs associated with the three methods on this new dataset division ($\mathcal{X}^2(2)=141.11$, $p<.001$). Using pairwise Wilcoxon signed ranks tests with Bonferroni p-value adjustments, we found that our tracing network performed significantly better (\textbf{\emph{MAP} = .834}) than VSM (\textbf{\emph{MAP} = .625}; $p<.001$) and LSI (\textbf{\emph{MAP} = .637}; $p<.001$). As such, we address RQ2 and conclude that in general our tracing network  improved trace link accuracy in comparison to both VSM and LSI, and that improvements were more marked as the size of the training set increased. We expect additional improvements by reconfiguring the tracing network for use with a larger training set \cite{bengio2012practical}.
However, we also observed that when using the original training set (i.e. 45\% of data), our tracing network only outperformed LSI at higher levels of recall as shown in Figure \ref{fig:prCurve_45_45_10} and Figure \ref{fig:PRCurve_before_after}. In future work, we plan to explore the trade-offs between these two methods for specific data features, and to further improve the performance of the tracing network. 




\section{Related Work}
\label{sec:relatedWork}
In this section we focus on prior work that has integrated ontology, semantics, or NLP into the tracing process.  Researchers have attempted to improve bag-of-word approaches such as the Vector Space Model (VSM) \cite{DBLP:journals/tse/HayesDS06} by integrating matching terms, project glossaries, and other forms of thesauri \cite{DBLP:journals/tse/HayesDS06}. Basic enhancements have included user feedback techniques such as Rocchio \cite{DBLP:journals/tse/HayesDS06} or Direct Query manipulation (DQM) \cite{DBLP:conf/sac/ShinC12} to increase or decrease term weights. However, these approaches fail to leverage semantic information.

Other techniques identify terms for briding the term mismatch between source and target artifacts.  Dietrich et al. utilized validated trace links to identify frequent item sets of terms occurring across pairs of source and target artifacts, and then used these to augment the text in the trace query \cite{DBLP:conf/kbse/DietrichCS13}.   Gibiec et al. \cite{Marek:ASE2010} approached trace query augmentation by acquiring related documents from the Internet, and then extracting domain related terms. Researchers have also used phrase detection and chunking to search for requirements impacted by change requests \cite{DBLP:conf/esem/AroraSBZG13} or to improve the trace retrieval process \cite{DBLP:journals/ese/ZouSC10}. None of these techniques attempted to understand semantics of the artifacts.

Researchers have also explored the use of knowledge bases to create and utilize semantically aware associations in the trace creation process -- where a knowledge base include basic domain terms and sentences that describe the relationships between those terms \cite{JacksonP},\cite{DBLP:reference/db/Gruber09},\cite{DBLP:journals/tkde/ShvaikoE13}. Data is typically represented as an ontology in which relationships are represented using \emph{AND}, \emph{OR}, \emph{implication}, and \emph{negation} operators \cite{JacksonP}. Traceability researchers have proposed the idea of using ontology to connect source and target artifacts \cite{conf/apsec/HayashiYS10, DBLP:journals/kais/AssawamekinSP10}. Approaches have been proposed for weighting the evidence for a trace link according to distance between concepts in the ontology \cite{Yonghua}. Unfortunately, building domain-specific ontologies is time consuming and ontologies are generally not available for technical software engineering domains.

Finally, while researchers have proposed techniques that more closely mimic the way human analysts reason about trace links and perform tracing tasks \cite{DBLP:conf/re/0004CB13, DBLP:conf/iwpc/MahmoudNX12}, there is very limited work in this area. Our prior work with DoCIT, described in Section \ref{sec:intro}, is one exception  \cite{DBLP:conf/re/0004CB13}. DoCIT utilizes both ontology and heuristics to reason over concepts in the domain in order to deliver accurate trace links.  However, as previously explained, DoCIT requires non-trivial setup costs to build a customized ontology and heuristics for each domain, and is sensitive to flaws in syntactic parsing.  In contrast, the RNN approach described in this paper requires only a corpus of domain documents and a training set of validated trace links. 

On the other hand, deep learning has been successfully applied to many software engineering tasks. For example, Lam et al. combined deep neural networks with information retrieval techniques to identify buggy files in bug reports  \cite{lam2015combining}. Wang et al. utilized a deep belief network to extract semantic features from source code for the purpose of defect prediction \cite{wang2016automatically}. Raychev et al. adopted RNN and N-gram to build the language model for the task of synthesizing code completions \cite{raychev2014code}. We were not able to find work that applied deep learning techniques to traceability tasks in our literature review.  
\section{Threats to Validity}
Two primary threats to validity potentially impact our work.  First, due to the challenge of obtaining large industrial datasets including artifacts and trace links, and the time needed to experiment with different algorithms for learning word embeddings and generating trace links, our work focused on a single domain of Positive Train Control. As a result, we cannot claim generalizability.  However, the PTC dataset included text taken from external regulations, and written by multiple requirements engineers, systems engineers, and developers. The threat to validity arises primarily from the possibility that characteristics of our specific dataset may have impacted results of our experiment. For example, the size of the overall dataset, the characteristics of the vocabulary used, and/or the nature of each individual artifact, may make our approach more or less effective. In the next phase of our work we will evaluate our approach on additional datasets. 

Second, we cannot guarantee that the trace matrix used for evaluation is 100\% correct.  However, it was provided by our industrial collaborators and used throughout their project to demonstrate coverage or regulatory codes. The metrics we used (i.e. MAP, Recall, and Precision) are all accepted research standards for evaluating trace results \cite{DBLP:journals/tse/HayesDS06}.  To avoid comparison against a weak baseline, we report comparisons against two standard baselines: VSM and LSI and configured them using a Genetic Algorithm. 


\section{Conclusions}
\label{sec:conclusion}
In this paper, we have proposed a neural network architecture that utilizes word embedding and RNN techniques to automatically generate trace links. The Bidirectional Recurrent Gated Unit effectively constructed semantic associations between artifacts, and delivered significantly higher MAP scores than either VSM or LSI when evaluated on our large industrial dataset. It also notably increased both precision and recall. Given an initial training set of trace links, our tracing network is fully automated and highly scalable. In  future work, we will focus on improving precision of the tracing network by identifying and including more representative negative examples in the training set.  

The tracing network is currently trained to process natural language text. In future work, we will investigate techniques for applying it to other types of artifacts such as source code or formatted data. Finally, given the difficulty and limitations of acquiring large corpora of data we will investigate hybrid approaches that combine human knowledge with the neural network. In summary, the findings we have presented in this paper have demonstrated that deep learning techniques can be effectively applied to the tracing process. We see this as a non-trivial advance in our goal of automating the creation of accurate trace links in industrial-strength datasets. 
\section{Acknowledgments}
The work in this paper was partially funded by the US National Science Foundation Grant CCF-1319680.

\vspace{10pt}
\bibliographystyle{abbrv}

\end{document}